\documentclass[twocolumn,nofootinbib,prd,floatfix,preprintnumbers,amsmath,amssymb,groupedaddress,superscriptaddress]{revtex4}
\usepackage{epsfig}  
\usepackage{array}  
\usepackage{hyperref}

\newcommand{\AdS}{\mathrm{AdS}}
\newcommand{\Lie}{\,\mathrm{Lie}} 
\newcommand{\tr}{\,\mathrm{tr}}
\newcommand{\str}{\,\mathrm{str}} 
\begin{document}   
\title{Anomaly Holography, the Wess-Zumino-Witten Term, and Electroweak Symmetry Breaking} 
\author{Ben Gripaios}  
\email{ben.gripaios@epfl.ch}  
\affiliation{Institut des Th\'{e}ories des Ph\'{e}nom\`{e}nes Physiques, EPFL, CH-1015 Lausanne, Switzerland.}
\affiliation{CERN, Theory Division, CH-1211 Geneva 23, Switzerland.}
\begin{abstract} 
I consider anomalies in effective field theories (EFTs) of gauge fields coupled to fermions on an interval in $AdS_5$, and their holographic duals.
The anomalies give rise to constraints on the consistent EFT description, which are stronger than the usual four-dimensional anomaly cancellation condition for the zero modes.
Even though the anomalies occur on both boundaries of the interval, corresponding to both the UV and the IR of the holographic dual, they are nevertheless consistent with the non-renormalization of the anomaly and the
't Hooft matching condition.
They give rise, in general,  to a Wess-Zumino-Witten (WZW) term in the four-dimensional, low-energy effective action, whose form I compute.
Finally I discuss the relevance to holographic models of electroweak symmetry breaking. I show that the so-called `minimal composite Higgs models' have a consistent EFT description without a WZW term.
In contrast, a variant of an earlier model of Contino, Nomura, and Pomarol does have a WZW term.
\end{abstract}   
\maketitle 
\section{\label{intro}Introduction}
The AdS/CFT \cite{Maldacena:1997re} correspondence has given rise to new solutions to the hierarchy problem of electroweak symmetry breaking (EWSB) \cite{Randall:1999ee}.
In their most sophisticated form \cite{Contino:2003ve,Agashe:2004rs}, these consist of gauge theories coupled to fermions on an interval (or orbifold) of an $AdS_5$ geometry, with the scalar Higgs sector of the Standard Model arising from the fifth-dimensional components of the gauge fields. Such models 
are dual \cite{Arkani-Hamed:2000ds,Rattazzi:2000hs} to models in four dimensions, in which a CFT, coupled to external fields in the ultra-violet, becomes strongly coupled in the infra-red. The onset of strong coupling spontaneously breaks some of the symmetries of the theory. In particular, the breaking of approximate global symmetries of the theory gives rise to pseudo-Nambu-Goldstone bosons in the low-energy effective theory, that play the r\^{o}le of light Higgs scalars and give rise to EWSB.\footnote{Although I will not discuss them here, similar considerations apply to Higgsless models \cite{Csaki:2003sh} and models with  other, {\em e.\ g.\ }flat, geometries.}

In general, gauge theories of this type, living on an interval in $d=5$, suffer from anomalies. These anomalies, which are localized on the four-dimensional boundaries of the interval, arise variously from
fermions (whether localized on the boundaries or propagating in the bulk) and from bulk Chern-Simons (CS) terms for the gauge fields. This has been known for some time \cite{Arkani-Hamed:2001is,Scrucca:2004jn}, but appears to have been disregarded in the literature on holographic models of EWSB, perhaps with the tacit assumption that, provided the usual $d=4$ anomaly cancellation condition is satisfed for the gauge and fermion zero modes (which of course it is in any model that reproduces the Standard Model at low energies), the fermionic boundary anomalies can always be cancelled by suitable CS terms. 

Unfortunately, this is not the case. Indeed, consider \cite{Gripaios:2007tk} the example of an $SU(2)$ bulk gauge field, broken to the same $U(1)$ subgroup on each boundary,
and let the fermionic content consist of a left-handed Weyl fermion of charge $+1$ on one boundary, and a left-handed Weyl fermion of  charge $-1$ on the other boundary.
Now, the theory is certainly free of anomalies from the $d=4$, low-energy perspective, because its zero modes describe a $U(1)$ gauge theory coupled to Weyl fermions of charge $\pm1$.
But there is no way in which one can cancel the boundary anomalies in the $d=5$ theory with a CS term. The reason for this is simply that the CS term, which is proportional to $\str T^A T^B T^C \equiv \tr T^A \{T^B, T^C\}$, vanishes identically for the generators $T^A$ of the Lie algebra of $SU(2)$. 

The observation that this simple example illustrates, namely that $d=4$ anomaly cancellation is necessary but not sufficient for $d=5$ anomaly cancellation, is not new. To my knowledge, it appeared first in the context of string orbifold models in \cite{vonGersdorff:2003dt} (and later in \cite{Liao:2006uja}). Nevertheless, it does not seem to be widely appreciated.

It is clear, then, that one needs to worry about the anomalies in holographic models of EWSB.
Since such theories are non-renormalizable, effective field theories, the anomalies do not render the theory inconsistent. Rather, similar to the case of anomalous EFTs in $d=4$ \cite{Preskill:1990fr}, a consistent EFT description always exists, but the anomalous symmetries must be non-linearly realized on the boundaries in $d=5$ \cite{Gripaios:2007tk}. The effect of this non-linear realization is to change the boundary conditions (BCs) of the gauge fields, such that the spectrum 
of gauge field zero modes is not that which one might na\"{\i}vely assume. It was further explained in \cite{Gripaios:2007tk} how, in this consistent description, the usual anomaly cancellation condition is guaranteed in the $d=4$ dual. Essentially, what happens is that any boundary anomaly leads to the corresponding gauge field being non-linearly realized, such that it cannot be present in the low-energy gauge group. In consequence, the surviving low-energy gauge group is reduced to one that is anomaly-free with respect to the $d=4$ zero modes. Moreover, in this description, the 't Hooft matching condition \cite{tHooft:1980xb} is obeyed in the $d=4$ dual \cite{Gripaios:2007tk}.

Conversely, to get the pattern of symmetry breaking that is na\"{\i}vely assumed in a given model, the net anomaly for the unbroken subgroup on the boundary must vanish,
and it is important that this be checked in existing and future models. What is more, even if the anomalies can be cancelled, there will still, in general, be observable physics associated with the anomaly, that one might hope to explore at upcoming collider experiments. This physics takes the form of  Wess-Zumino-Witten (WZW) terms \cite{Wess:1971yu,Witten:1983tw} in the low-energy effective action,
that reproduce the UV anomalies of non-linearly realized symmetries, in accordance with 't Hooft's condition \cite{Panico:2007qd}. This is analogous to what happens in the strong interaction. There, the WZW term in the low-energy pion Lagrangian reproduces the anomalies of the approximate chiral symmetries of the quarks in the UV, and gives rise to spectacular physics, most notably the decay $\pi^0 \rightarrow \gamma \gamma$. In the context of holographic models of EWSB, the electromagnetic field is enlarged to the full Standard Model gauge group, and the the Higgs sector takes the place of the pions.

The goal of the present work is to supply a strategy by which one can firstly determine if a given model can be made anomaly-free by adding CS terms, and secondly can be used to compute any WZW term in the low-energy effective action. Explicitly, the strategy is to rewrite the theory in terms of an equivalent theory  in which the full bulk gauge symmetry is resurrected on the boundaries (albeit non-linearly realized), via the addition of coset sigma models (which I review in the next Section) localized on the boundaries. In Section \ref{class},
I show that this always possible at the classical level. 
In Section \ref{cosa}, I consider quantum effects, beginning with a review of anomalies in coset sigma models.
In Section \ref{quant}, I return to theories on an interval of $AdS_5$, and determine whether $G$-invariance may be resurrected at the quantum level, including the anomalies.
In doing so, I derive the consistency conditions, and the form of the WZW term that arises in the low-energy effective Lagrangian. 
In Section \ref{ex}, I apply the strategy. I show that the so-called `$MCHM_{10}$' model \cite{Agashe:2006at} can be made consistent by addition of CS terms, and does not have any WZW term in the low energy Lagrangian. Nor does the  `$MCHM_{5}$' model, who consistency was shown previously in \cite{Gripaios:2007tk}. I show that yet another model, derived from the one of \cite{Contino:2003ve}, does have a WZW term.

I should remark that WZW terms are generic in models in which the Higgs sector arises from anomalous, non-linearly-realized symmetries. For example, Hill and Hill \cite{Hill:2007nz} have recently discussed how they arise in little Higgs models. 

I postpone a discussion of the phenomenological implications to future work, but for two remarks. The first is that the WZW terms may prove to be invaluable if Nature does actually choose to realize EWSB via strong-coupling at the TeV scale.
The reason for this is that excited states in such theories typically have a width comparable to their mass (there is no small parameter to suppress one versus the other)
and so the spectrum contains a mess of broad resonances, whose identification in experiment is problematic. This makes it difficult to learn anything about the theory from experiment,
  even before one attacks the theoretical strong-coupling problem. Again, the analogy with the strong interaction is helpful: after many decades of experiment at the GeV scale, we still argue about the spectrum of hadronic resonances and our understanding is minimal, even though we know very well what the microscopic theory is in that case. The things that we can infer about the high-energy theory from low-energy experiments (such as the presence of approximate chiral symmetries of quarks and the number of colours) come from the study of the symmetries and anomalies thereof. If a similar scenario does indeed explain the weak scale, then it is on the symmetries and their anomalies that we should perhaps focus.
  
  The second remark is that WZW terms should be of particular interest in models, such as the ones discussed here, where the Standard Model fermions themselves belong partly to the strongly-coupled sector. In such models, the WZW term, which measures the anomaly content of the strongly-coupled sector, is fixed by the Standard Model fermion content, once it has been decided how the Standard Model fermions fit into representations of the symmetry group of the strongly-coupled sector. This offers the hope that we may be able to {\em predict} the form of the WZW term.
  
Our notations for $\AdS$ and fermions therein are those of \cite{Gripaios:2006dc}.
\section{\label{cos}$G/H$-coset sigma models}
In this Section, I review the construction of coset sigma models, and the gauging thereof. I follow closely the notation and discussion of Preskill \cite{Preskill:1990fr}.
Let $G$ be a compact Lie group with subgroup $H$, and let $G/H$ be a coset manifold with co-ordinates $\phi_a$. I can always choose a reductive splitting for a basis of Hermitian generators, $T^\alpha \in \Lie (H)$, and the remaining generators, $X^a \in \Lie (G)$, such that 
\begin{gather} \label{reduct}
[T^\alpha,X^a] = i f^{ \alpha a}_{\phantom{ \alpha a}b} X^b 
\end{gather}
and I shall always do so in the sequel. A coset representative, $g(\phi_a) \in G$, is defined up to the equivalence relation
\begin{gather}
g \sim gh, 
\end{gather}
where $h \in H$;
a convenient coset representative is
\begin{gather}
g(\phi_a) = e^{i \phi_a X^a}.
\end{gather}
Under left-multiplication by $\Omega^{-1} \in G$, the coset representative transforms as
\begin{gather} \label{Gaction}
\Omega: \;  g(\phi_a) \rightarrow g(\phi^{\prime}_a) = \Omega^{-1} g(\phi_a) h (\Omega, g),
\end{gather}
where the compensating transformation, $h (\Omega, g)$, is chosen so as to maintain the choice of coset representative.
The $\Lie (G)$-valued Cartan $1$-form,\footnote{The reason for the obscure notation $0^g$ will, I hope, become clear.} $0^g \equiv g^{-1} d g$, transforms under the global $G$-action (\ref{Gaction}) as
\begin{gather}
\Omega:  \;0^g \rightarrow  h^{-1}   g^{-1} \Omega d( \Omega^{-1} g h  )= h^{-1}  0^g h   + h^{-1}  d h.
\end{gather}
Decomposing the Cartan form as
\begin{gather}
0^g = (0^g)_H + (0^g)_X = (0^g)_\alpha T^\alpha + (0^g)_a X^a,
\end{gather}
the reductive splitting (\ref{reduct}) implies that the global $G$ action (\ref{Gaction}) reduces to
\begin{align}
\Omega: \; (0^g)_H &\rightarrow  h^{-1}  (0^g)_H h   + h^{-1}  d h  \\
(0^g)_X &\rightarrow h^{-1}  (0^g)_X h .
\end{align}
Thus, $(0^g)_X$ transforms homogeneously under the global $G$-action (\ref{Gaction}), whereas $(0^g)_H$ transforms as an $H$-connection.

I should now like to elevate the global $G$ action (\ref{Gaction}) to a local one. To do so, I let $\Omega = \Omega (x)$, and define the local $G$-action 
on the coset representative by 
\begin{gather} \label{Glocal}
\Omega:  \; g(\phi_a) \rightarrow g(\phi^{\prime}_a) = \Omega^{-1} (x) g(\phi_a) h (\Omega (x), g).
\end{gather}
I also introduce a connection for $G$, {\em viz.} a $\Lie (G)$-valued 1-form, $A$, which I take to be anti-hermitian, transforming as 
\begin{gather}
\Omega:  \;A \rightarrow A^{\Omega} \equiv \Omega^{-1} (A + d) \Omega.
\end{gather}
Then, the object $A^g \equiv g^{-1} (A + d) g$ ({\em cf.}\ the definition of $0^g  \equiv g^{-1} d g$) transforms under the local $G$-action as
\begin{gather}
\Omega: \; A^g \rightarrow A^{\Omega \Omega^{-1} gh} = h^{-1} A^g h + h^{-1} d h.
\end{gather}
Decomposing $A^g \in \Lie(G)$ as before, one sees again that $(A^g)_X$ transforms homogeneously under the local $G$-action (\ref{Glocal}), whereas $(A^g)_H$ transforms as an $H$-connection.

Given some matter field $\psi$ transforming as a representation $r$ of $H$, 
\begin{gather}
\psi \rightarrow D_r [h^{-1}] \psi,
\end{gather}
this can be extended to a transformation under the $G$-action (global or local) as
\begin{gather}\label{psireal}
\Omega:  \;\psi \rightarrow D_r [h^{-1}(\Omega, g)] \psi.
\end{gather}
Note that this transformation, which involves the coset fields, is non-linear.
A covariant derivative for the local $G$-action is
\begin{gather}\label{A}
D\psi = (d + D_r [(A^g)_H ] ) \psi,
\end{gather}
(or
$D\psi = (d + D_r [(0^g)_H ] ) \psi $
in the global case).

I will also need to consider a matter field $\Psi$, transforming as a representation of the whole group $G$,
\begin{gather}
\Omega:  \;\Psi \rightarrow D_R [\Omega^{-1}] \Psi.
\end{gather}
Whilst such fields can be trivially coupled to the $G$-connection $A$ using the usual covariant derivative, they can also be coupled to matter fields $\psi$ transforming linearly under $H$, but non-linearly under $G$, as discussed above. Indeed, the field $\Psi^\prime \equiv D_R [g^{-1}]\Psi$ transforms only under the compensator
\begin{gather} \label{bigreal}
\Omega: \; \Psi^\prime \rightarrow D_R [h^{-1}(\Omega, g)] \Psi^\prime
\end{gather}
and can be coupled to fields $\psi$ transforming as in (\ref{psireal}). This is the non-linear sigma model analogue of a Yukawa coupling.

We now have all the necessary ingredients to built a locally $G$-invariant, $G/H$-coset sigma model, coupled to matter transforming in representations
of either $H$ or $G$. 
\section{\label{class}Resurrecting $G$-invariance at the classical level}
In the usual formulation of a gauge theory on an interval, the bulk $G$ gauge invariance is allowed to be broken to subgroups 
$H_{0,1}$ on the boundaries, by choosing the Dirichlet BC for the fifth components, $A_5$, of gauge fields corresponding to generators in $\Lie (H_0)$ on the UV boundary, and the Neumann BC for the others. Similarly, on the IR boundary, one chooses the Dirichlet BC for the $A_5$ components corresponding to generators in $\Lie (H_1)$, and so on. Since the theory on the boundary does not respect the full $G$ invariance, matter fields living on the boundary need only come in representations of the subroup $H_0$ on the UV boundary, and $H_1$ on the IR boundary. Similarly, though matter fields that propagate in the bulk must transform as representations of $G$, their boundary conditions need only respect $H_0$ or $H_1$, as appropriate. In particular, for a bulk (Dirac) fermion, one is free to choose either the left- or right-handed Weyl components to vanish on, say, the UV boundary, provided that states with the same BC furnish a representation of $H_0$.

In this Section, I show that, at least at the classical level, such a theory has an equivalent formulation in which the full, bulk $G$-invariance is maintained everywhere, including on the boundaries. In this formulation, I must add $G/H_{0,1}$-coset sigma models (as described in the last Section) on the respective boundaries. The symmetries corresponding to generators in $G/H_{0,1}$ are non-linearly realized by the coset scalar fields. The couplings on the boundaries between the coset scalars and the gauge fields modify the BCs and give rise to the same physical spectrum of gauge boson zero modes as in the usual formulation. A pedagogical explanation is given in {\em e.\ g.\ }\cite{Csaki:2005vy}.

So the full, $G$ gauge invariance can be resurrected, at least in the gauge sector, by adding $G/H_{0,1}$-coset sigma models on the boundaries. What is more, the same coset fields can be used to restore the gauge symmetry in the matter sector as well.

To see how this is achieved, consider first matter fields localized on, say, the UV boundary. In the usual formulation, these need only transform as a rep of the broken subgroup $H_0$. But as we saw in the previous Section, the $G/H_{0}$ coset fields allow us to extend the matter field to a realization of $G$, according to (\ref{psireal}).

Resurrecting $G$-invariance for a matter field $\Psi$ living in the bulk is not much more difficult. Here, the problem is that, in the usual formalism, the BCs for $\Psi$ on, say, the UV boundary, need not respect $G$, but only the subgroup $H_0$. Let us suppose, for example, that we have a bulk fermion
\begin{gather}
\Psi = \begin{pmatrix} \psi_{\alpha} \\ \overline{\chi}^{\dot{\alpha}}  \end{pmatrix}
\end{gather} 
in a rep $R$ of $G$, and that the UV BCs are $\psi_0 = 0$ for states in $R$ forming some rep $r$ of $H_0$, and $\chi_0 =0$ otherwise.
To resurrect $G$ on the UV boundary, I consider instead a bulk fermion $\Psi$ with the $G$-invariant BC $\chi_0 = 0$ for all states in the rep $R$. I also add a boundary-localized fermion $\eta_0$ in rep $\overline{r}$ of $H_0$. Now $\eta$ carries a realization of $G$ according to (\ref{psireal}) and, furthermore, the object $D_R[g^{-1}]\psi$ transforms only under the compensator, 
as in (\ref{bigreal}). Since the rep $R$, construed as a rep of $H_0$, contains the rep $r$, I can write a $G$-invariant term coupling $\eta_0$ and $D_R[g^{-1}]\psi_0$ on the UV boundary.
In the limit that the dimensionful coupling constant of this term becomes large (of order of the EFT cut-off), its effect \cite{Gripaios:2006dc} is equivalent to flipping the BC from $\chi_0 = 0$ to $\psi_0 = 0$ for states in a rep $r$. Thus, it is equivalent to the usual situation of a bulk fermion with BCs respecting only $H_0$.

On the IR boundary, I follow a similar procedure, except that I choose the $G$-invariant BC for $\Psi$ to be the opposite one, namely $\psi_1 = 0$, for all states. With this choice, the bulk fermion $\Psi$ has no $d=4$ zero modes. This simplifies the derivation of the low-energy effective action: to get it, I simply integrate out all of the (massive) bulk fermion modes.

We thus see how to convert the usual formulation, with $G$ broken on the boundaries, into one with $G$-invariance resurrected on the boundaries. The equivalence of these two formulations is, in fact, a trivial one: the alternative formulation simply has a larger gauge invariance than the usual one, in that it is $G$-invariant everywhere. The usual formulation is then obtained as a gauge-fixing of the alternative one. The gauge-fixing is, of course, the one in which the coset fields on the boundaries vanish.

Given that the two formulations are equivalent, what is the utility of the alternative formulation? As we shall see, it makes it much easier to deduce the consistency requirements following from anomaly considerations, and also to compute the WZW term.

Thus far, everything has been classical. In the next section, we shall see how things change at the quantum level.
\section{\label{cosa}Anomalies in $G/H$-coset sigma models}
In this Section I review the anomaly structure of coset models, following Alvarez-Gaum\'{e} and Ginsparg \cite{AlvarezGaume:1985yb}.
Let us first recall the structure of the anomaly in $d=4$ arising from a Weyl fermion, $\psi$, transforming as a linear representation, $R$ of group $G$. The effective action, obtained by integrating out the fermions, and defined by\footnote{In treating the theory at the quantum level, I shall always consider the Euclidean path integral formalism.}
\begin{gather}
e^{-\Gamma_R [A]} = \int_{\psi} \exp{-\int d^4 x \; \mathcal{L} (A,\psi)},
\end{gather} 
is not, in general, invariant under an infinitesimal gauge transformation, $A\rightarrow A^{1+\omega}$, but rather transforms as
\begin{gather}\label{standanom}
\delta_{\omega} \Gamma_R [A] = \frac{1}{24 \pi^2}  \int d^4 x \; Q_R (\omega , A),
\end{gather} 
where
\begin{gather}
Q_R (\omega , A) = \tr_R (\omega d[AdA +\frac{1}{2}A^3])
\end{gather}
with the trace over matrices in the representation $R$.
To integrate this, I let $\omega = g^{-1} \delta g$, such that
\begin{gather}
 \Gamma_R [A^{g+\delta g}] -  \Gamma_R [A^g] = \frac{1}{24 \pi^2}  \int d^4 x \; Q_R ( g^{-1} \delta g , A^g),
\end{gather}
and choose a one-parameter family $g_s (x)$ of maps on $s \in [0,1]$, such that $g_{s=0} = 1$ and $g_{s=1} = g$. Integrating with respect to $s$, I obtain
\begin{gather}\label{end}
\Gamma_R [A^g] - \Gamma_R [A] = \frac{1}{24 \pi^2} \int_{0}^{1} ds \; \int d^4 x \; Q_R ( g_s^{-1} \partial_s g_s , A^{g_s}).
\end{gather}

Now consider fermions in a representation $r$ of the subgroup $H$, coupled to the gauge field and the sigma-model fields via the $H$-connection, $(A^g)_H$, as in (\ref{A}). By comparison with (\ref{standanom}) we see that the $G$-anomaly of the effective action $\Gamma_r$, obtained by integrating over the fermions, is given by
\begin{gather} \label{ranom}
\delta_{\epsilon} \Gamma_r [(A^g)_H] = \frac{1}{24 \pi^2}  \int d^4 x \; Q_r (\epsilon , (A^g)_H).
\end{gather}
In the above, $\epsilon$ is the infinitesimal version of the compensating $h$ transformation: $h (\Omega, g) = 1 + \epsilon (\Omega, g)+\dots$.

In theories on a higher-dimensional interval, with $G$-resurrected at the classical level as described in the previous Section, the boundary fermions will give rise to anomalies of exactly this form, with $H$ replaced by the linearly-realized subgroup $H_{0,1}$ on the relevant boundary. In order to consistently quantize the theory, I need to be able to cancel this anomaly against anomalies coming from the bulk fermions and CS terms. The latter are anomalies of the group $G$, whereas the anomalies in (\ref{ranom}) have the structure of anomalies in the subgroup $H = H_{0,1}$ (even though they are defined for the whole group $G$ via the compensator). It would, therefore, seem to be impossible to cancel the anomalies in this way. 

In fact, the anomalies can be cancelled, under certain conditions. To see how this may be achieved, consider the following object
\begin{gather}
\Gamma_R^{WZW} = \frac{1}{24 \pi^2}  \int_0^1 ds \; \int d^4 x \; Q_R (g_s \partial_s g_s^{-1} , (A^g)_H^{g_s^{-1}}),
\end{gather}
where $R$ is any representation of $G$. By reversing the argument of Eqn's (\ref{standanom}-\ref{end}), we see that this object transforms under the $G$-action like the difference\footnote{Note that $(A^g)_H^{g^{-1}} \neq (A)_H$!}
\begin{gather}\label{diff}
\Gamma_R [(A^g)_H^{g^{-1}}] - \Gamma_R [(A^g)_H].
\end{gather}
These terms are just the effective actions one would obtain by integrating out fermions in representation $R$, coupled to connections $(A^g)_H^{g^{-1}}$ and $(A^g)_H$,
respectively. But under the $G$-action, $(A^g)_H \rightarrow (A^g)_H^h$, where $h = h(\Omega,g)$ is the compensator, and so the anomalous $G$-action on the second term in (\ref{diff}) cancels the anomalous $G$-action on $\Gamma_r [(A^g)_H]$ in (\ref{ranom}) iff.\ 
\begin{gather}\label{match}
\str_r T^\alpha T^\beta T^\gamma = \str_R T^\alpha T^\beta T^\gamma,
\end{gather}
where the generators are those of $\Lie (H)$.
Moreover, since under the $G$-action $(A^g)_H^{g^{-1}} \rightarrow (A^g)_H^{g^{-1}\Omega}$, we see that the first term in (\ref{diff}) has the usual $G$-anomaly corresponding to representation $R$, but with the alternative $G$-connection, $(A^g)_H^{g^{-1}}$, replacing the usual $G$-connection $A$. This is easily corrected by addition of Bardeen's counterterm \cite{Bardeen:1969md,AlvarezGaume:1984dr}
\begin{widetext}
\begin{gather}
B_R [A_1, A_2] = \frac{1}{48\pi^2} \int d^4 x \; \tr_R [(F_1 + F_2)(A_2 A_1 - A_1 A_2) - A_2^3  A_1+ A_1^3 A_2 +\frac{1}{2} A_2 A_1 A_2 A_1],
\end{gather}
which transforms such that
\begin{gather}
B_R [A_1^\Omega, A_2^\Omega] - B_R[A_1, A_2] = \Gamma_R [A_1^\Omega] - \Gamma_R [A_2^\Omega] - \Gamma_R [A_1] + \Gamma_R [A_2].
\end{gather}
\end{widetext}
In the case at hand, setting $A_1 = A$ and $A_2 = (A^g)_H^{g^{-1}}$, I find that adding the term
\begin{gather} \label{whole}
\Gamma_R^{WZW} + B_R [A, (A^g)_H^{g^{-1}} ]
\end{gather}
to the action converts the $G$-anomaly of a fermion in representation $r$ of $H$ to the usual $G$-anomaly of a fermion in representation $R$ of $G$, iff.\ the $H$-anomalies of $r$ and $R$ match, in the sense of (\ref{match}). If this is the case, then I can cancel the anomalies coming from boundary fermions against anomalies coming from bulk fermions or CS terms.

\section{\label{quant}Resurrecting $G$-invariance at the quantum level}
In order to resurrect $G$-invariance everywhere on a $d=5$ interval at the quantum level, the $G$ anomalies on each of the two boundaries must separately vanish: If they do not, the number of linearly-realized gauge symmetries (and hence the low-energy gauge group) is smaller than that which is claimed. On each boundary, there are three contributions to the anomaly. Firstly, there are boundary-localized fermions in a reps $r_{0,1}$ of $H_{0,1}$, whose contribution to the anomaly takes the form of (\ref{ranom}), with $r\rightarrow r_{0,1}$ and $H \rightarrow H_{0,1}$. Secondly, there are bulk fermions in a rep $R'$ of $G$. Thirdly, there are CS terms corresponding to a rep $R''$ of $G$. The nature of the BCs I choose for the bulk fermions means that the contribution to the anomaly is the same for both bulk fermions and CS terms. They take the form of (\ref{standanom}) with $R = R' \oplus R''$, but have opposite signs on the two boundaries. Equivalently, I can say that the anomaly on the UV boundary is that of $R$, whilst the anomaly on the IR boundary is that of $\overline{R}$. 

Now, we saw in the last Section that anomalies of the form (\ref{ranom}), can be converted to anomalies of the form (\ref{standanom}), via the term (\ref{whole}) iff.\ (\ref{match}) is satisfied. Therefore, on the interval, we can consistently quantize the theory with the assumed structure of linearly and non-linearly realized symmetries iff.\ the anomaly of the rep
$r_0$ of $H_0$ matches that of the rep $R$, construed as a rep of $H_0$, {\em and}  the anomaly of the rep
$r_1$ of $H_1$ matches that of the rep $\overline{R}$, construed as a rep of $H_1$. 

Note that this condition includes the usual $d=4$ zero mode anomaly cancellation condition, which is that the anomaly of the rep $r_0 \oplus \overline{r}_1$ of the largest subgroup of both $H_0$ and  $H_1$ should vanish, but it is in fact much stronger. What is more, even if I allow myself free choice of the CS term, corresponding to $R$ being an arbitrary rep, I still find that the condition is stronger than the usual $d=4$ condition. We see, in particular, that our original example, with $SU(2)$ broken to $U(1)$ on each boundary and with fermions of opposite charge on the boundaries, does not satisfy the condition, because the anomaly of any rep $R$ of $SU(2)$ must vanish. In this example, the condition is not satisfied at either boundary. To exhibit an example where the condition is satisfied at one boundary but not the other, consider the same set-up, but with the bulk group $SU(2)$ completely broken on one boundary. The condition is now trivially satisfied on this boundary, and from the $d=4$ perspective (there is no surviving gauge group in either case), but is violated on the other boundary, where a $U(1)$ is preserved.

Once a consistent theory has been found, it is a simple matter to derive the form of the WZW term in the low-energy effective action. Since I have a theory which is everywhere $G$-invariant, I am free to choose the gauge $A_5 = 0$. When I do this, the bulk Dirac fermions have no chiral coupling, and integrating them out has no effect on the anomaly structure. Furthermore, any CS term involves only vector fields in this gauge, so cannot contribute to the WZW term. The only place the low-energy WZW term can now come from is from the boundary-localized WZW terms of the form (\ref{whole}). What is more, I can use some of the remaining gauge freedom to gauge away all of the coset fields on one boundary, such that the WZW term on that boundary vanishes.
Once I have done so, I can still gauge away some, but not all, of the coset fields on the other boundary.The ones I can gauge away are those that were not paired with a coset field on the other boundary, in the sense that they corresponded to the same generator of $G$. Thus I am left with one physical scalar coset field in the low-energy theory for every generator that is in $\Lie (G)$ but not in $\Lie (H_0)$ or in $\Lie (H_1)$.

The WZW term that remains on one boundary in this gauge is not quite the WZW term that appears in the low-energy effective action in $d=4$, because it involves all of the $G$ gauge fields. Some of these gauge fields do not survive in the low-energy theory, that is to say they have no zero modes. The only ones that do survive are those corresponding to generators in the intersection of $\Lie(H_0)$ and $\Lie(H_1)$. To integrate out the massive gauge fields in the WZW term, I simply set them to zero, and replace the surviving gauge fields by their zero modes. Having done so, I am left with the WZW term that appears in the low-energy effective action.

Though it may at first seem rather odd that I can evaluate the WZW term in the low-energy action either by going to a gauge in which it is generated at the UV boundary, or by going to a gauge in which it is generated at the IR boundary, this is in fact completely necessary from the point of view of the holographic dual. According to the duality, the UV boundary corresponds to the UV of the $d=4$ theory, and the IR boundary to the IR. Because the anomaly is non-renormalized, its form is the same at any energy scale. So the form of the anomaly in the $d=4$ dual is completely fixed by the anomalies on, say, the UV boundary. 

Although the form of the anomaly is fixed, the form of the WZW term that appears in the low-energy effective action is not.  Indeed, the form of the WZW term is not fixed until the fate of the various symmetries at low energy has been decided: If, on the one hand, a symmetry remains linearly-realized, the anomaly must be reproduced by fermions in the low-energy effective theory, as argued by 't Hooft; if, on the other hand, the symmetry is non-linearly realized, then the anomaly is reproduced at low energy by the WZW term. Now, in theories on an interval in $AdS$, the fate of the symmetries at low energies is decided, in part, by the anomaly structure on the IR boundary: if a symmetry is anomalous on the IR boundary, it must be non-linearly realized at low energy.

So in the context of holography, one can say, in a sense, that the anomaly in the $d=4$ dual is completely determined by the anomaly on the UV boundary of the $d=5$ theory, but that the WZW term in the $d=4$ dual is then determined by the anomaly on the IR boundary in $d=5$.
\section{\label{ex}Examples}
Perhaps the most realistic holographic models of EWSB are the `minimal composite Higgs models' of \cite{Agashe:2004rs}. They are based on bulk gauge group $G = SU(3)_c \times SO(5) \times U(1)_X$, broken to the custodially-symmetric $H_1 = SU(3)\times SO(4) \times U(1)_X$ in the IR, where $SO(4) = SU(2)_L \times SU(2)_R$, and to the Standard Model gauge group, $H_1 = SU(3)\times SU(2)_L \times U(1)_Y$ in the IR, where $Y=T_R^3 + X$. The custodial symmetry prevents large corrections to the $T$-parameter of precision tests of EWSB, and by enlarging the $SO(4)$ to $O(4)$ one can even control the corrections to $Z \rightarrow b\overline{b}$ \cite{Agashe:2006at}. The models still appear to require some fine-tuning to get a small enough value for the $S$-parameter, however.

It is simple enough to see that models based on this pattern of symmetry breaking can always be rendered consistent by addition of a suitable CS term, and do not lead to a WZW term in the low-energy effective action. Indeed, in the alternative formulation with $G$ invariance resurrected everywhere, we know that the boundary-localized fermions correspond to the fermion zero modes, which in this case are just three Standard Model generations. Now, in general, the fermions can be split between the two boundaries, with some living on the UV boundary and some living on the IR boundary. But in the case at hand, all of the zero mode fermions must live on the UV boundary, where the gauge group is that of the Standard Model. If some of the fermions were to live on the IR boundary, then we would have to be able to organize them into a representation of the unbroken group there, {\em viz.} $H_1 = SU(3)\times SO(4) \times U(1)_X$.
But there simply is no way to organize a subset of the Standard Model fermions into a rep of $SU(3)\times SO(4) \times U(1)_X$. Thus all of the Standard Model fermions live on the UV boundary, and the net $H_1$ anomaly from fermions localized on the UV boundary vanishes. Of course, there will in general be be anomalies coming from bulk fermions, but these come in reps of $G$ and can always be cancelled by a CS term. There is, therefore, no WZW term in the low-energy effective action.

It is not difficult to find a model which does have a WZW term. Indeed, consider the symmetry structure of the original holographic model of this type, in which bulk group $G =SU(3)_c \times SU(3)_L \times U(1)_X$ is broken on both branes to $H_{0,1} = SU(3)_c \times SU(2)_L \times U(1)_Y$, where $SU(2)_L$ is generated by the first three Gell-Mann generators of $SU(3)$ and $Y = T^8/\sqrt{3}+X$.\footnote{This model has many faults, not least that it is inconsistent with electroweak precision tests without fine-tuning, but that does not concern us here.} Again, in the formulation with $G$ resurrected everywhere, the boundary fermions must correspond to the Standard Model fermions. But now I am free to put some fermions, the quarks say, on one boundary, and the leptons on the other. The boundary fermion contributions to the  $SU(2)_L^2 U(1)_Y$ and $U(1)_Y^3$ anomalies are now non-vanishing, and must be cancelled by a combination of bulk CS terms and boundary WZW terms. The boundary WZW terms give rise to a WZW term in the low-energy effective action, as described above.
\begin{acknowledgments}
I thank S. M. West for providing details of his calculations of the fermion anomalies in various models, and thank
G.~ F.~ Giudice, L.~ Randall, R.~ Rattazzi and A.~ Wulzer for discussions.
\end{acknowledgments}

\end{document}